\documentclass[a4paper,11pt]{article}
\usepackage{jheppub}
\usepackage[T1]{fontenc}
\usepackage{amsmath}
\usepackage{mathrsfs}
\newcommand{\dd}{\mathrm d}
\newcommand{\ii}{\mathrm i}
\newcommand{\Dop}{\mathscr D}
\newcommand{\adj}{\operatorname{adj}}
\newcommand{\tr}{\operatorname{tr}}

\allowdisplaybreaks[2]

\title{\boldmath Exact algebraic separation of the gravitoelectromagnetic perturbations of Kerr-Newman-(anti-)de Sitter black holes}
\author[]{Jun-Kun Zhao}
\affiliation[]{Institute of Theoretical Physics, Chinese Academy of Sciences, Beijing 100190, China}
\emailAdd{junkunzhao@itp.ac.cn}

\abstract{We obtain an explicit separation of the coupled gravitational and electromagnetic radiative equations on nonextremal Kerr--Newman--(anti-)de Sitter black holes with nonzero charge and rotation. By directly matching the Fourier coefficients, we construct a separating matrix with constant determinant that is independent of the frequency and azimuthal number and applies to both spin sectors. We establish that this transformation is smooth and invertible throughout the open AdS exterior and between the event and cosmological horizons in dS. The resulting radial and angular systems are valid at arbitrary complex frequencies. We also translate the regularity of the radiative Newman--Penrose variables at the rotation axes and future horizons into boundary conditions for the separated solutions. In the asymptotically flat limit, the transformation is related to Hintz's matrix integrating factor by the allowed frame freedom. These results provide a basis for studying quasinormal modes and black hole stability in Kerr--Newman--(anti-)de Sitter spacetimes, as well as exploring holographic applications to strongly coupled systems.}
\arxivnumber{2610.22274}

\begin{document}
\maketitle
\flushbottom
\section{Introduction}
Linear perturbations of black holes encode rich information about their dynamical properties, such as linear stability and quasinormal-mode (QNM) spectra~\cite{Berti:2009kk}. Characterized by complex frequencies, QNMs describe the oscillations and decay of perturbations~\cite{Vishveshwara:1970zz}, providing valuable insights from both theoretical and observational perspectives. In particular, they govern the ringdown phase of gravitational waves following black hole mergers~\cite{LIGOScientific:2016aoc,LIGOScientific:2016lio}, while perturbation analyses can also reveal dynamical instabilities, including those associated with superradiance under suitable conditions~\cite{Brito:2015oca}. For black holes in anti-de Sitter (AdS) spacetime, linear perturbations acquire an additional significance through the AdS/CFT correspondence~\cite{Maldacena:1997re,Ammon:2015wua}. An AdS black hole is holographically dual to a thermal state of a conformal field theory~\cite{Witten:1998zw}, and its perturbations characterize the response of the dual strongly coupled system to small disturbances. In this framework, black hole QNMs correspond to poles of retarded correlation functions in the dual field theory~\cite{Son:2002sd,Kovtun:2005ev}. Their complex frequencies encode the response of perturbations, with the imaginary parts determining characteristic relaxation timescales as the system approaches thermal equilibrium~\cite{Horowitz:1999jd,Berti:2009kk,Cardoso:2013pza}.

The study of linear perturbations of stationary, rotating black holes has long been a central topic in black hole physics. A major breakthrough was achieved by Teukolsky, who exploited the Petrov type-D structure of the Kerr spacetime to obtain a unified master equation for massless fields of different spins, including scalar, electromagnetic, and gravitational perturbations~\cite{Teukolsky:1972my,Teukolsky:1973ha}. This equation admits separation into radial and angular ordinary differential equations, providing a powerful framework for investigating black hole dynamics, QNM spectra, and stability. This separated structure also played a crucial role in Whiting's proof of mode stability~\cite{Whiting:1988vc}. Although the Kerr--Newman (KN) spacetime is also of Petrov type D, its nonzero background Maxwell field couples gravitational and electromagnetic perturbations at linear order, preventing a straightforward extension of Teukolsky's separation procedure. Consequently, despite extensive analytical and numerical investigations, an exact separation of the coupled gravitoelectromagnetic radiative equations had remained elusive for decades~\cite{Chitre:1976bb,Chandrasekhar1983,KalninsWilliams1998,Dudley:1977zz,Berti:2005eb,Pani:2013wsa,Dias:2015wqa,Giorgi:2020ujd,Dias:2022oqm}.

Recently, Hintz made a significant advance by introducing a matrix integrating factor that achieves exact separation of the coupled radiative equations for asymptotically flat KN black holes~\cite{Hintz:2026jiw}. Subsequent computations of QNM frequencies based on these separated equations showed excellent agreement with independent numerical results, providing strong support for the construction~\cite{Destounis:2026ljd}. The separated system has since been used to obtain charge-regular radiative amplitudes and tidal Love numbers~\cite{Motohashi:2026wac,Chakraborty:2026kyf}. Extending this separation to black holes in asymptotically AdS and dS spacetimes is nontrivial, since both the perturbation equations and the spacetime boundary structure are modified~\cite{Dias:2013sdc,Hintz:2016jak}. Motivated by these advances, we investigate whether Hintz's separation approach can be generalized to KN black holes with a cosmological constant.

Here we construct an explicit matrix transformation that separates the coupled gravitoelectromagnetic radiative equations on nonextremal Kerr--Newman--(anti-)de Sitter (KN--(A)dS) black holes with nonzero rotation and charge. Independent of the frequency and azimuthal number, the matrix applies to both positive- and negative-spin sectors, yielding radial and angular systems valid for arbitrary complex frequencies. We establish its smooth invertibility outside the event horizon in AdS and between the event and cosmological horizons in dS. We also derive regularity conditions for the radiative Newman--Penrose (NP) variables at the rotation axes and future horizons, and relate our transformation to Hintz's matrix integrating factor in the asymptotically flat limit.

The remainder of this paper is organized as follows. Section~\ref{sec:setup} introduces the background and perturbation equations, while Sec.~\ref{sec:separation} presents the exact separation. Section~\ref{sec:ds} extends the construction to dS, and Sec.~\ref{sec:points} examines axis and horizon regularity. Section~\ref{sec:discussion} concludes with a discussion of the results and open questions. The appendices provide supporting derivations, regularity proofs, and the asymptotically flat comparison.

\section{Background and perturbation equations} \label{sec:setup}
We consider the Kerr–Newman–AdS black hole in geometric units and with signature $(+---)$. The line element and gauge potential are given by
\begin{equation}
\begin{split}
\dd s^2= &\frac{\Delta}{\Sigma} \left(\dd t-\frac{a\sin^2\theta}{\Xi}\dd\phi\right)^2 -\frac{\Sigma}{\Delta}\dd r^2 -\frac{\Sigma}{T^2}\dd\theta^2 -\frac{T^2\sin^2\theta}{\Sigma} \left[a\,\dd t-\frac{r^2+a^2}{\Xi}\dd\phi\right]^2 \,,\\
\mathcal A= &-\frac{Qr}{\Sigma} \left(\dd t-\frac{a\sin^2\theta}{\Xi}\dd\phi\right) \,,
\end{split}
\end{equation}
where
\begin{equation}
\begin{split}
\Delta&=(r^2+a^2)\left(1+\frac{r^2}{L^2}\right)-2Mr+Q^2 \,,\qquad
\Sigma =r^2+a^2\cos^2\theta \,,\\ \Xi &=1-\frac{a^2}{L^2} \,,\qquad
T=\sqrt{1-\frac{a^2}{L^2}\cos^2\theta} \,.
\end{split}
\end{equation}
Here $L$ is the AdS radius, with $\Lambda=-3/L^2$, and $M$, $a$ and $Q$ are the mass, rotation and charge parameters. The limit $L\to\infty$ gives the asymptotically flat Kerr--Newman solution~\cite{Newman:1965my}. We take $M>0$, $Q\ne0$ and $0<|a|<L$, and consider a nonextremal black hole with outer horizon $r=r_+>0$. We work outside the horizon, where $r>r_+$ and $\Delta>0$. The de Sitter case is discussed in Sec.~\ref{sec:ds}.

We write $f=r+\ii a\cos\theta$ and $g=\bar f$, so that $fg=\Sigma$. In coordinates $(t,r,\theta,\phi)$, the generalized Kinnersley tetrad~\cite{Kraniotis:2018zmh} has components
\begin{equation}
\begin{split} \label{eq:tetrad}
\ell^\mu&=\frac1\Delta \left(r^2+a^2, \Delta, 0, a\Xi\right)  \,,\\
n^\mu&=\frac1{2\Sigma} \left(r^2+a^2, -\Delta, 0, a\Xi\right) \,,\\
m^\mu&=\frac1{\sqrt2f} \left( \ii \frac{ a\sin\theta}{T}, 0, T, \ii \frac{\Xi}{T\sin\theta}\right) \,.
\end{split}
\end{equation}
The null tetrad is normalized by $\ell^\mu n_\mu=1$ and $m^\mu\bar m_\mu=-1$. We fix the NP conventions~\cite{Newman:1961qr} by writing $e_1=\ell^\mu\partial_\mu$, $e_2=n^\mu\partial_\mu$, $e_3=m^\mu\partial_\mu$, and $e_4=\bar m^\mu\partial_\mu$. The components of the NP spin connection are defined by $\gamma^{\mathrm{NP}}_{ijk}=g_{\mu\nu} e_j^\nu e_k^\lambda (\nabla_\lambda e_i^\mu)$, with $\gamma^{\mathrm{NP}}_{ijk}=-\gamma^{\mathrm{NP}}_{jik}$ by metric compatibility. In particular, $\kappa_{\mathrm{NP}}=-\gamma^{\mathrm{NP}}_{311}$ and $\sigma_{\mathrm{NP}}=-\gamma^{\mathrm{NP}}_{313}$.

We use the curvature convention $[\nabla_\mu, \nabla_\nu]v^\alpha = R^\alpha{}_{\beta\mu\nu} v^\beta$. The Weyl scalars are defined by $\Psi_0=-C_{1313}, \Psi_1=-C_{1213}, \Psi_2=-C_{1342}, \Psi_3=-C_{1242}, \Psi_4=-C_{2424}$. For the electromagnetic field $F_{\mu\nu}=\partial_\mu\mathcal A_\nu-\partial_\nu\mathcal A_\mu$, the Maxwell scalars are $ \Phi_0=F_{13}, \Phi_1=\tfrac12(F_{12}+F_{43}), \Phi_2=F_{42}$. Here $C_{ijkl}$ and $F_{ij}$ denote the components of the Weyl tensor and electromagnetic field in the null frame.

For the KN--AdS background and the null tetrad~\eqref{eq:tetrad}, the associated nonvanishing NP spin coefficients are
\begin{equation}
\begin{split}
\rho_{\mathrm{NP}} &=-\frac{1}{\bar f} \,,\qquad   \mu_{\mathrm{NP}}=-\frac{\Delta}{2f\bar f^{\,2}} \,,\qquad
\pi_{\mathrm{NP}} =\frac{\mathrm{i}aT\sin\theta} {\sqrt{2}\bar f^{\,2}} \,,\\
\tau_{\mathrm{NP}} &=-\frac{\mathrm{i}aT\sin\theta} {\sqrt{2}\Sigma} \,,\qquad \beta_{\mathrm{NP}}=\frac{T\cot\theta+ \partial_\theta T}{2\sqrt{2}f} \,,\\
\alpha_{\mathrm{NP}} &=\pi_{\mathrm{NP}} -\overline{\beta_{\mathrm{NP}}} \,,\qquad  \gamma_{\mathrm{NP}}=\mu_{\mathrm{NP}}+\frac{\partial_r \Delta}{4\Sigma} \,.
\end{split}
\end{equation}
The remaining coefficients satisfy $\kappa_{\mathrm{NP}}=\sigma_{\mathrm{NP}} =\lambda_{\mathrm{NP}}=\nu_{\mathrm{NP}}=\varepsilon_{\mathrm{NP}}=0$. The background Weyl and Maxwell scalars are
\begin{equation}
\Psi_2=\frac{Q^2-Mf}{f\bar f^{\,3}} \,,\quad \Phi_1=\frac{Q}{2\bar f^{\,2}} \,,
\end{equation}
with $\Psi_0=\Psi_1=\Psi_3=\Psi_4=0$ and $\Phi_0=\Phi_2=0$.

We now consider linear gravitational and electromagnetic perturbations of this background, denoting first-order perturbations by a superscript $(1)$. Since the background Maxwell scalar $\Phi_1$ is nonzero for $Q\ne0$, we use the infinitesimal null-rotation freedom of the tetrad to impose the phantom gauge, $\Phi_0^{(1)}=\Phi_2^{(1)}=0$.

In the NP formalism, the radiative perturbations are described by two sets of coupled first-order equations, one involving fields of positive spin weights $+1,+2$ and the other involving fields of negative spin weights $-1,-2$~\cite{Chandrasekhar1983}. Under $m^\mu\mapsto e^{\ii\chi}m^\mu$ and $\bar m^\mu\mapsto e^{-\ii\chi}\bar m^\mu$, with $\ell^\mu,n^\mu$ fixed and $\chi$ real, a component of weight $w$ acquires the phase $e^{\ii w\chi}$. In particular, the perturbed Weyl scalars $\Psi_0^{(1)},\Psi_1^{(1)}$ have spin weights $+2,+1$, while $\Psi_4^{(1)},\Psi_3^{(1)}$ have spin weights $-2,-1$. We will use a single invertible matrix $X$ to separate the positive- and negative-spin systems. Following~\cite{Hintz:2026jiw}, the positive-spin variables are
\begin{equation}
\psi_2 =\Psi_0^{(1)} \,, \quad \psi_1 =\sqrt2g\Psi_1^{(1)} \,, \quad
k =\frac{\kappa_1}{\sqrt2g^2} \,, \quad s=\frac{f\sigma_1}{g^2} \,.
\end{equation}
Here $\kappa_1=\kappa_{\mathrm{NP}}^{(1)}$ and $\sigma_1=\sigma_{\mathrm{NP}}^{(1)}$ denote the linearized NP spin coefficients. For the negative-spin sector, we define
\begin{equation}\label{eq:negative-variables}
\psi_{-2}=g^4\Psi_4^{(1)} \,,\quad \psi_{-1}=\frac{g^3}{\sqrt2}\Psi_3^{(1)} \,,\quad  k_-=\frac{fg}{\sqrt2}\nu_1 \,,\quad  s_-=\frac g2\lambda_1 \,,
\end{equation}
where $\nu_1=\nu_{\mathrm{NP}}^{(1)}$ and $\lambda_1=\lambda_{\mathrm{NP}}^{(1)}$. The two quadruples $(\psi_2,\psi_1,k,s)$ and $(\psi_{-2},\psi_{-1},k_-,s_-)$ have spin weights $(2,1,1,2)$ and $(-2,-1,-1,-2)$, respectively.

Stationarity and axisymmetry allow us to consider modes proportional to $e^{-\ii\omega t+\ii m\phi}$, with $\omega\in\mathbb C$ and $m\in\mathbb Z$. For such a mode, we define the radial and angular operators by
\begin{equation}\label{eq:frequencies}
\mathcal D_j=\partial_r+\frac{\ii K}{\Delta} +j\frac{\partial_r \Delta}{\Delta} \,,\qquad \mathcal L_j=T\partial_\theta+\mathcal U +j(T\cot\theta+ \partial_\theta T) \,,
\end{equation}
where $K=a\Xi m-(r^2+a^2)\omega$ and $\mathcal U=(\Xi m\csc\theta-a\omega\sin\theta)/T$. The daggered operators $\mathcal D_j^\dagger$ and $\mathcal L_j^\dagger$ are obtained from $\mathcal D_j$ and $\mathcal L_j$ by replacing $(\omega,m)$ with $(-\omega,-m)$, without complex conjugation.

Following the phantom-gauge treatment of asymptotically flat KN perturbations in~\cite{Chandrasekhar1983}, we carry out the corresponding reduction on the KN--AdS background. Combining the linearized Bianchi and Ricci identities with the source-free Maxwell equations, and using the variables and differential operators defined above, we obtain the following coupled first-order system for the positive-spin sector:
\begin{equation}
\begin{split} \label{eq:np4}
\left(\mathcal L_2-\frac{3F_0}{g}\right)\psi_2
-\left(\mathcal D_0+\frac{3}{g}\right)\psi_1 &=-2kF_+   \,,\\
\Delta\left(\mathcal D_2^\dagger-\frac{3}{g}\right)\psi_2
+\left(\mathcal L_{-1}^\dagger+\frac{3F_0}{g}\right)\psi_1 &=2sF_-  \,,\\
\left(\mathcal D_0+\frac{3}{g}\right)s
-\left(\mathcal L_{-1}^\dagger+\frac{3F_0}{g}\right)k &=\frac{f\psi_2}{g^2} \,,\\
\Delta\left(\mathcal D_2^\dagger-\frac{3}{g}\right)k
+\left(\mathcal L_2-\frac{3F_0}{g}\right)s &=\frac{2f\psi_1}{g^2} \,,
\end{split}
\end{equation}
where $F_0=\ii aT\sin\theta$ and $F_\pm=3(M-Q^2/f)\pm Q^2g/f^2$. In the limit $L\to\infty$, this system reduces to the first-order system for asymptotically flat KN perturbations~\cite{Chandrasekhar1983,Hintz:2026jiw}.

The negative-spin equations can be related to the positive-spin system~\eqref{eq:np4} by defining the rescaled fields $\widehat\psi_2 =\Delta^{-2}\psi_{-2}, \widehat\psi_1=-\Delta^{-1}\psi_{-1}, \widehat k=-\Delta^{-2}k_-, \widehat s=\Delta^{-1}s_-$. These fields satisfy~\eqref{eq:np4} with $(\omega,m)$ replaced by $(-\omega,-m)$. We exploit this relation to treat the two sectors simultaneously below.

\section{Exact separation}\label{sec:separation}
Following~\cite{Hintz:2026jiw}, we label the positive- and negative-spin sectors by $\varsigma=+1$ and $-1$, respectively, and define
\begin{subequations} \label{eq:chiralvars}
\begin{equation}
 y_\pm^{(+)}=\Delta^{5/4}\sqrt{T\sin\theta}
 \begin{pmatrix}
  \psi_2\pm\ii\Delta^{-1/2}\psi_1\\
  -2Q(k\pm\ii\Delta^{-1/2}s)
 \end{pmatrix} \,,
\end{equation}
\begin{equation}
 y_\pm^{(-)}=\Delta^{-3/4}\sqrt{T\sin\theta}
 \begin{pmatrix}
  \psi_{-2}\mp\ii\sqrt\Delta\,\psi_{-1}\\
  2Q(k_-\mp\ii\sqrt\Delta\,s_-)
 \end{pmatrix} \,.
\end{equation}
\end{subequations}
The superscript labels the spin sector, while the subscript $\pm$ labels the two chiral combinations within each sector. In terms of these variables, the four coupled equations in each spin sector take the compact form
\begin{equation}
\begin{split}\label{eq:chiral}
(\Dop-A)y_+^{(\varsigma)}   &=(c^{(\varsigma)}I-B^{(\varsigma)})y_-^{(\varsigma)}  \,,\\
(\bar\Dop+A)y_-^{(\varsigma)}  &=(c^{(\varsigma)}I+B^{(\varsigma)})y_+^{(\varsigma)} \,,
\end{split}
\end{equation}
where $\Dop=\sqrt\Delta\,\partial_r-\ii T\partial_\theta$, $\bar\Dop=\sqrt\Delta\,\partial_r+\ii T\partial_\theta$, and $I$ is the $2\times2$ identity matrix. The coefficients are
\begin{equation}
\begin{split}
 A&=\begin{pmatrix}
       0&-\dfrac{\ii Qg}{f^2}\\
       \dfrac{\ii Qf}{g^2}&0
     \end{pmatrix} \,,\qquad
 B^{(\varsigma)} =\begin{pmatrix}
       \ii b^{(\varsigma)}&\dfrac{3\ii(Mf-Q^2)}{Qf}\\
       \dfrac{3\ii Qf}{g^2}&-\ii b^{(\varsigma)}
     \end{pmatrix} \,,\\
b^{(\varsigma)}&=-\varsigma\mathcal U -\frac32(T\cot\theta+\partial_\theta T)+\frac{3F_0}{g} \,,\qquad
c^{(\varsigma)} =\frac{\ii\varsigma K}{\sqrt\Delta} +\frac{3\sqrt\Delta}{g}-\frac{3\partial_r\Delta}{4\sqrt\Delta} \,.
\end{split}
\end{equation}
Only the terms proportional to $\omega,m$ depend on $\varsigma$. The matrix $A$ is the same in both sectors and is Hermitian, with $\tr A=\tr B^{(\varsigma)}=0$. The cosmological constant enters through the background functions $\Delta,T,\Xi$.

We apply the same field transformation in both sectors,
\begin{equation}  \label{eq:change}
z_+^{(\varsigma)}=(\det X) X^{-1}y_+^{(\varsigma)} \,,\qquad z_-^{(\varsigma)}=X^\dagger y_-^{(\varsigma)} \,.
\end{equation}
Here $X(r,\theta)$ is invertible and independent of $\omega,m,\varsigma$, and $X^\dagger$ denotes its Hermitian conjugate. In the transformed $z_+^{(\varsigma)}$ equation, the matrix multiplying $z_-^{(\varsigma)}$ is
\begin{equation}  \label{eq:transformedcoupling}
C^{(\varsigma)} =(\det X) X^{-1}(c^{(\varsigma)}I-B^{(\varsigma)})(X^\dagger)^{-1} \,.
\end{equation}
Motivated by Hintz's separation of perturbations of the asymptotically flat Kerr--Newman black hole~\cite{Hintz:2026jiw}, we seek a matrix $X$ for which $C^{(\varsigma)}$ takes the following separated form in the Pauli basis $\{I,\sigma_1,\sigma_2,\sigma_3\}$,
\begin{equation}\label{eq:separatedC}
 C^{(\varsigma)} = R_0^{(\varsigma)}(r)I +\ii R_1^{(\varsigma)}(r)\sigma_2 +\ii\Theta_1^{(\varsigma)}(\theta)\sigma_1
 +\ii\Theta_0^{(\varsigma)}(\theta)\sigma_3 \,,
\end{equation}
where $\sigma_j$ are the Pauli matrices. The radial part of $C^{(\varsigma)}$ commutes with $\sigma_2$, while its angular part anticommutes with $\sigma_2$,
\begin{equation}
\bigl[\sigma_2,\, R_0^{(\varsigma)}I+\ii R_1^{(\varsigma)}\sigma_2\bigr] = 0 \,, \qquad
\bigl\{\sigma_2 \,, \ii\Theta_1^{(\varsigma)}\sigma_1 +\ii\Theta_0^{(\varsigma)}\sigma_3\bigr\} = 0 \,.
\end{equation}
In the eigenbasis of $\sigma_2$, the radial terms are thus diagonal and the angular terms are off-diagonal. This structure guides the construction of $X$ below and underlies the separation ansatz in~\eqref{eq:ansatz}.

The algebraic separation condition does not fix the transformation matrix uniquely. The separated form~\eqref{eq:separatedC} is preserved in both sectors by
\begin{equation}  \label{eq:freedom}
 X\longmapsto XE \,,\qquad E=s_r(r)s_\theta(\theta)e^{u(r)\sigma_2}e^{\ii v(\theta)\sigma_2} \,,
\end{equation}
where $s_r,s_\theta$ are positive smooth functions and $u,v$ are real smooth functions, all independent of $\omega,m,\varsigma$. The factors $E$ form an Abelian group. After division by the positive scalar factor $s_r s_\theta$, each matrix belongs to $SO(2,\mathbb C)$. The matrix $C^{(\varsigma)}$ transforms as
\begin{equation}
C^{(\varsigma)}\longmapsto (\det E)E^{-1}C^{(\varsigma)}(E^\dagger)^{-1} \,.
\end{equation}
The radial exponential mixes the coefficients of $I$ and $\sigma_2$, while the angular exponential rotates those of $\sigma_1$ and $\sigma_3$. Below we use this freedom to make $\det X$ constant.

\subsection{An explicit choice of \texorpdfstring{$X$}{X}}
We now present an explicit choice of $X$ and the corresponding four coefficients of $C^{(\varsigma)}$. In Appendix~\ref{app:algebraic-checks}, we provide a detailed construction of $X$ by matching the Fourier coefficients and show how to recover it from the eigendirections of a matrix satisfying a transport equation.\footnote{See recent approaches to the separation of coupled perturbations based on transported eigendirections~\cite{Gray:2026rzu}, or on a Gram-matrix criterion and transport conditions~\cite{Furugori:2026pjx}.}
Define the positive constants
\begin{equation} \label{eq:alg-parameters}
\kappa=\sqrt{\left[M^2-Q^2\left(1+\frac{a^2}{L^2}\right)\right]^2 -\frac{4a^2Q^4}{L^2}} \,, \quad \eta=\frac{2Q^2} {M^2-Q^2(1+a^2/L^2)+\kappa} \,.
\end{equation}
Using the scalar freedom in~\eqref{eq:freedom}, we choose the normalized representative
\begin{equation}  \label{eq:alg-frame}
X=\frac1{\kappa Q\sqrt{A_rD_\theta}}
 \begin{pmatrix}
  \dfrac{\ii MaA_r\cos\theta-\nu D_\theta}{f}
   &\dfrac{\ii Q(aTA_r\sin\theta-D_\theta\sqrt{\Delta})}{f}\\[6pt]
  -\dfrac{\ii Q(aTA_r\sin\theta+D_\theta\sqrt{\Delta})}{g}
   &\dfrac{\ii MaA_r\cos\theta+\nu D_\theta}{g}
 \end{pmatrix}  \,.
\end{equation}
Here $A_r=r^2-\eta a^2$, $D_\theta=a^2(\cos^2\theta+\eta)$, and $\nu=Mr-Q^2$. The scalar factor $(A_rD_\theta)^{-1/2}$ leaves $C^{(\varsigma)}$ unchanged and gives the constant determinant
\begin{equation}
\det X=-1/(\kappa Q^2) \,.
\end{equation}
For the rotating, charged, nonextremal KN--AdS background considered here, $X$ and the field transformation~\eqref{eq:change} are smooth and invertible throughout $r>r_+$, $0<\theta<\pi$; see Appendix~\ref{app:invertibility}. Appendix~\ref{app:flat-equivalence} relates $X$ to Hintz's matrix integrating factor~\cite{Hintz:2026jiw} in the limit $L\to\infty$. For this choice, $C^{(\varsigma)}$ takes the form~\eqref{eq:separatedC}.
\begin{subequations}
\label{eq:alg-separated-coefficients}
Explicitly, the radial coefficients are
\begin{align}
R_0^{(\varsigma)} &= \frac{3\partial_r\Delta}{4\sqrt\Delta} -\frac{3r\sqrt\Delta}{A_r} -\frac{\ii\varsigma K}{\sqrt\Delta} +\frac{2\ii\varsigma aQ^2\sqrt\Delta}{\kappa A_r} [a(1+\eta)\omega-\Xi m] \,,\\
 \ii R_1^{(\varsigma)} &= -\frac{3[M(r^2+\eta a^2)-2Q^2r]}{2QA_r} +\frac{2\ii\varsigma aQ\nu}{\kappa A_r}   [a(1+\eta)\omega-\Xi m] \,,
\end{align}
and the angular coefficients are
\begin{align}
\Theta_1^{(\varsigma)} &= \frac{3Ma^2(\cos^2\theta-\eta)}{2QD_\theta} +\frac{2\varsigma a^2MQ\cos\theta}{\kappa D_\theta} [a(1+\eta)\omega-\Xi m] \,,\\
 \Theta_0^{(\varsigma)} &= -\frac32(T\cot\theta+\partial_\theta T) -\frac{3a^2T\sin\theta\cos\theta}{D_\theta} -\varsigma\mathcal U \notag\\
 &\qquad -\frac{2\varsigma a^2Q^2T\sin\theta}{\kappa D_\theta} [a(1+\eta)\omega-\Xi m] \,.
\end{align}
\end{subequations}
Here $K$ and $\mathcal U$ are defined below Eq.~\eqref{eq:frequencies}.

\subsection{Transformed perturbation equations}

Applying~\eqref{eq:change} with the normalized matrix~\eqref{eq:alg-frame} to~\eqref{eq:chiral} gives
\begin{equation}
\begin{split} \label{eq:alg-chiral}
[\Dop+(h_r+h_\theta)\sigma_2]z_+^{(\varsigma)} &=C^{(\varsigma)}z_-^{(\varsigma)} \,,\\
[\bar\Dop-(h_r+h_\theta)\sigma_2]z_-^{(\varsigma)} &=\adj C^{(\varsigma)}z_+^{(\varsigma)} \,,
\end{split}
\end{equation}
where
\begin{equation}
h_r=\frac{M\eta a^2-Q^2r}{QA_r} \,,\qquad h_\theta=\frac{M\eta a^2}{QD_\theta} \,.
\end{equation}
They are real and the same in both spin sectors, with $h_r$ depending only on $r$ and $h_\theta$ only on $\theta$.

The above equations are obtained with a constant-determinant normalization of $X$, which simplifies the derivative terms but is not essential for separation. As shown in Appendix~\ref{app:algebraic-checks}, the scalar rescaling $\widetilde X=s_r(r)s_\theta(\theta)X$ allowed by Eq.~\eqref{eq:freedom} gives
\begin{equation}
\widetilde X^{-1}(\Dop\widetilde X-A\widetilde X) =(h_r+h_\theta)\sigma_2 +\left(\sqrt{\Delta}\,\partial_r\log s_r-\ii T\,\partial_\theta\log s_\theta\right)I \,.
\end{equation}
The corresponding transformed fields $\widetilde z_\pm^{(\varsigma)}=s_rs_\theta z_\pm^{(\varsigma)}$ satisfy
\begin{equation}
\begin{split}
\left[\Dop+(h_r+h_\theta)\sigma_2 -\left(\sqrt{\Delta}\,\partial_r\log s_r-\ii T\,\partial_\theta\log s_\theta\right)I\right]\widetilde z_+^{(\varsigma)} &=C^{(\varsigma)}\widetilde z_-^{(\varsigma)} \,,\\
\left[\overline{\Dop}-(h_r+h_\theta)\sigma_2 -\left(\sqrt{\Delta}\,\partial_r\log s_r+\ii T\,\partial_\theta\log s_\theta\right)I\right]\widetilde z_-^{(\varsigma)} &=\adj C^{(\varsigma)}\widetilde z_+^{(\varsigma)} \,.
\end{split}
\end{equation}
The additional identity terms separate into radial and angular parts, while $C^{(\varsigma)}$ remains unchanged. Thus, the perturbation equations can still be separated without fixing the scalar freedom in $X$. Since $A$ is traceless, the coefficient of $I$ in $\widetilde X^{-1}(\Dop\widetilde X-A\widetilde X)$ is given by $\Dop(\det\widetilde X)/(2\det\widetilde X)$ and hence vanishes for a constant determinant. This is the case for both our choice $\det X=-1/(\kappa Q^2)$ and the normalization $\det X=1$ adopted by Yoshida~\cite{Yoshida:2026yly}.

\subsection{Radial and angular equations}

To separate the fields in~\eqref{eq:alg-chiral}, write $\chi^{(\varsigma)}(\theta)=(\chi_1^{(\varsigma)}(\theta), \chi_2^{(\varsigma)}(\theta))^{\mathsf T}$ and introduce the ansatz
\begin{equation}  \label{eq:ansatz}
z_\pm^{(\varsigma)} =\tfrac12\big[u_1^{(\varsigma)}(r)I \pm\ii u_2^{(\varsigma)}(r)\sigma_2\big]\chi^{(\varsigma)}(\theta) \,.
\end{equation}
Substituting~\eqref{eq:ansatz} into~\eqref{eq:alg-chiral}, we add and subtract the two equations and separate the radial and angular dependence with a separation constant $\lambda_\varsigma$. Writing $\mathbf u^{(\varsigma)}= (u_1^{(\varsigma)}, u_2^{(\varsigma)}) ^{\mathsf T}$, the radial equations take the form
\begin{equation} \label{eq:radial}
\sqrt\Delta\,\partial_r\mathbf u^{(\varsigma)} =\big[ R_0^{(\varsigma)}\sigma_3 +R_1^{(\varsigma)}\sigma_1  +(h_r+\lambda_\varsigma)\sigma_2\big] \mathbf u^{(\varsigma)} \,.
\end{equation}
The corresponding angular equations are
\begin{equation}\label{eq:angular}
T\partial_\theta\chi^{(\varsigma)} =\big[ -\Theta_1^{(\varsigma)}\sigma_1 -\Theta_0^{(\varsigma)}\sigma_3  +\ii(\lambda_\varsigma-h_\theta)\sigma_2\big] \chi^{(\varsigma)} \,.
\end{equation}
In separating the equations, we used $\sigma_2^2=I$ and $\sigma_2\sigma_j=-\sigma_j\sigma_2$ for $j=1,3$.

The separated equations hold for complex $\omega$ in the region $r>r_+$. For either spin sector, combining radial and angular solutions with the same $\lambda_\varsigma$ in~\eqref{eq:ansatz} determines $z_\pm^{(\varsigma)}$. Inverting~\eqref{eq:change} and~\eqref{eq:chiralvars} then gives a solution of the original perturbation equations:~\eqref{eq:np4} for $\varsigma=+1$ and~\eqref{eq:negative-np4} for $\varsigma=-1$.

\section{Extension to de Sitter}\label{sec:ds}
The construction extends to KN--dS by replacing $L^{-2}$ with $-L^{-2}$ in the KN--AdS background functions,
\begin{equation}
\begin{split}
\Delta &= (r^2+a^2)\left(1-\frac{r^2}{L^2}\right)-2Mr+Q^2 \,, \\
T^2 &= 1+\frac{a^2}{L^2}\cos^2\theta \,, \quad \Xi = 1+\frac{a^2}{L^2} \,.
\end{split}
\end{equation}
The same replacement must be made in the parameters $\kappa$ and $\eta$. Here $L>0$ is the de Sitter radius, with $\Lambda=3/L^2$. We assume $M>0$, $aQ\neq0$, and consecutive positive simple event and cosmological horizons $0<r_b<r_c$, with $\Delta>0$ between them.

With these parameters, the algebraic identities underlying the separation remain valid. Consequently, the same expression for the normalized matrix $X$ yields the radial and angular equations for both spin sectors at arbitrary complex frequencies. As shown in Appendix~\ref{app:invertibility}, $X$ is smooth and invertible throughout the region between the event and cosmological horizons.

\section{Behavior at horizons and rotation axes}\label{sec:points}
Here, we examine the behavior of the separated solutions near the rotation axes and horizons for both spin sectors $\varsigma=\pm1$. We translate regularity of the radiative NP fields there into boundary conditions on the radial and angular functions.
\subsection{Rotation axes}
Near the north and south axes, we set $t=\theta$ and $t=\pi-\theta$, respectively, and define
\begin{equation}
 \alpha_N=\varsigma m+\frac32 \,,\qquad
 \alpha_S=\frac32-\varsigma m \,.
\end{equation}
Using $T(0)=T(\pi)=\sqrt{\Xi}$, the angular equation~\eqref{eq:angular} takes the asymptotic form
\begin{equation}
 \frac{\dd\chi^{(\varsigma)}}{\dd t} =\left[\frac{\alpha}{t}\sigma_3+O(1)\right] \chi^{(\varsigma)} \,,
\end{equation}
where $\alpha=\alpha_N$ or $\alpha_S$. The indicial exponents are $\pm\alpha$. The branch with exponent $-|\alpha|$ gives singular radiative NP variables and is excluded; the regular branch takes the form
\begin{equation}
 \chi^{(\varsigma)}=t^{|\alpha|}
 \begin{cases}
 \begin{pmatrix}
  1 & 0\\
  0 & t
 \end{pmatrix}
 \mathbf{c}(t^2) \,, & \alpha>0 \,,\\[6pt]
 \begin{pmatrix}
  t & 0\\
  0 & 1
 \end{pmatrix}
 \mathbf{c}(t^2) \,, & \alpha<0 \,,
 \end{cases} \label{eq:physical-axis-domain}
\end{equation}
where $\mathbf{c}=(c_1,c_2)^{\mathsf T}$, with $c_1,c_2$ smooth near $t=0$. As shown in Appendix~\ref{app:physical-boundaries}, this branch corresponds to smooth radiative NP variables.

Since $T\to\sqrt{\Xi}>0$ at both poles and $|\alpha|\geq\tfrac12$ for integer $m$, the singular branch behaves as $t^{-|\alpha|}$ and is not square-integrable with respect to $\dd\theta/T$. The divergence is logarithmic when $|\alpha|=\tfrac12$. Thus square integrability selects the regular branch in~\eqref{eq:physical-axis-domain}. For fixed real $\omega$, integer $m$, and $\varsigma$, the angular operator has a unique self-adjoint realization with a discrete real spectrum and a complete orthonormal eigenbasis in $L^2((0,\pi),\dd\theta/T;\mathbb C^2)$~\cite{Furugori:2026pjx}.

\subsection{Simple horizons}
At a simple horizon $r=r_h$, define $b_h^{(\varsigma)}=3/4-\ii\varsigma K(r_h)/\Delta'(r_h)$. Expanding the radial system near $r=r_h$, we obtain
\begin{equation}
\frac{d\mathbf u^{(\varsigma)}}{dr} =\left[\frac{b_h^{(\varsigma)}\sigma_3}{r-r_h} +O(|r-r_h|^{-1/2}) \right]\mathbf u^{(\varsigma)} \,.
\end{equation}
To specify the horizon conditions, define
\begin{equation}
H(r)=\exp\left(\ii\int^r\frac{K(s)}{\Delta(s)}\,ds\right) \,.
\end{equation}
The radial functions satisfy
\begin{equation}
\begin{split}\label{eq:physical-horizon-data}
u_1^{(\varsigma)} &\in H^{\zeta_h} \Delta^{1/4-\zeta_h\varsigma/2}C^\infty \,,\\
u_2^{(\varsigma)} &\in H^{\zeta_h} \Delta^{1/4-\zeta_h\varsigma}C^\infty \,,
\end{split}
\end{equation}
where $\zeta_h=+1$ for the ingoing condition at the future event horizon and $\zeta_h=-1$ for the outgoing condition at the future cosmological horizon. Here $C^\infty$ denotes functions that extend smoothly to $r=r_h$. The derivation is given in Appendix~\ref{app:physical-boundaries}.

\section{Discussion and outlook}\label{sec:discussion}
We have separated the coupled gravitoelectromagnetic radiative equations in the positive-spin ($\varsigma=+1$) and negative-spin ($\varsigma=-1$) sectors on nonextremal KN--(A)dS backgrounds with nonzero charge and rotation. The normalized matrix~\eqref{eq:alg-frame}, independent of $\omega$ and $m$, transforms the chiral system~\eqref{eq:chiral} into the radial and angular systems~\eqref{eq:radial} and~\eqref{eq:angular}, written uniformly for $\varsigma=\pm1$. The freedom~\eqref{eq:freedom} is used to set $\det X=-1/(\kappa Q^2)$ and remove the scalar derivative terms from both systems. Smooth invertibility holds outside the event horizon in AdS and between the event and cosmological horizons in dS.

We have also translated smoothness of the radiative NP variables at the rotation axes and future horizons into boundary conditions for the separated solutions. At the axes, these conditions select the regular branch of the angular system. In the asymptotically flat limit, our matrix $X$ is related to Hintz's integrating factor through the transformation~\eqref{eq:freedom}. For KN--AdS, the separated system provides a starting point for computing holographic QNMs at finite charge and rotation and for extracting the late-time relaxation timescales that govern thermalization in the dual field theory~\cite{Horowitz:1999jd,Son:2002sd,Kovtun:2005ev,Cardoso:2013pza}.

Several questions remain for future work. One is the reconstruction of metric and Maxwell perturbations from the separated radiative variables. Existing methods developed for Kerr~\cite{Merlin:2016boc,Berens:2024czo,Hollands:2024iqp} provide useful guidance, but their extension to KN–(A)dS must accommodate gravitational–electromagnetic coupling and variations of the black hole parameters. Another is the angular spectral problem at complex frequencies, where generalized eigenfunctions may be needed, as in the Kerr case~\cite{Finster:2015xma}. A further challenge is to impose gravitational and electromagnetic boundary conditions at the AdS boundary that are compatible with the field equations~\cite{Ishibashi:2004wx,Warnick:2013hba,Dias:2013sdc,Wald:2026irt}. Addressing these questions would facilitate systematic studies of quasinormal modes and superradiant instabilities in KN--AdS, as well as the radiative spectrum of KN--dS black holes.

\textit{Note added.} While preparing this manuscript, we became aware of related work by Yoshida~\cite{Yoshida:2026yly} on the separation of gravitoelectromagnetic perturbations of KN--dS black holes, which overlaps with part of our results.

\textit{Note added in v2.} Yoshida constructs the separating matrix using an auxiliary matrix that satisfies a transport equation, while in Appendix~\ref{app:X-matching} we obtain $X$ by matching the coefficients of $\omega$ and $\Xi m$. Gray and Ovcharenko subsequently extended the separation to Pleba\'nski--Demia\'nski spacetimes and derived a symmetry operator for the chiral system~\cite{Gray:2026rzu}. Furugori and Yoshida developed a general separation criterion for chiral systems and related their KN--dS separating matrix to ours through an explicit change of basis consistent with the frame freedom in Eq.~\eqref{eq:freedom}~\cite{Furugori:2026pjx}. Sarkar et al. presented an alternative separated formulation for KN--dS perturbations and applied it to the study of quasinormal modes and strong cosmic censorship~\cite{Sarkar:2026xkq}.

\acknowledgments
We thank Li Li and Tieguang Zi for helpful discussions. This work was supported by the National Natural Science Foundation of China under Grant No.~12505087.

\appendix
\section{Equations of the negative-spin sector}

In the same phantom gauge, the variables~\eqref{eq:negative-variables} satisfy
\begin{equation}
\begin{split} \label{eq:negative-np4}
 &\left(\mathcal L_2^\dagger-\frac{3F_0}{g}\right)\psi_{-2} +\Delta\left(\mathcal D_{-1}^\dagger+\frac3g\right)\psi_{-1} =2k_-F_+ \,,\\
 &\left(\mathcal D_0-\frac3g\right)\psi_{-2} -\left(\mathcal L_{-1}+\frac{3F_0}{g}\right)\psi_{-1} =2s_-F_- \,,\\
 &\Delta\left(\mathcal D_{-1}^\dagger+\frac3g\right)s_-  +\left(\mathcal L_{-1}+\frac{3F_0}{g}\right)k_-  =\frac{f\psi_{-2}}{g^2} \,,\\
 &\left(\mathcal D_0-\frac3g\right)k_- -\left(\mathcal L_2^\dagger-\frac{3F_0}{g}\right)s_-  =\frac{2f\psi_{-1}}{g^2} \,.
\end{split}
\end{equation}
Here $F_0,F_\pm,f,g$ retain their original definitions.

\section{Construction of the transformation matrix \texorpdfstring{$X$}{X}}\label{app:algebraic-checks}
As shown in Sec.~\ref{sec:setup}, the negative-spin system~\eqref{eq:negative-np4} reduces to the positive-spin system~\eqref{eq:np4} upon rescaling the fields and replacing $(\omega,m)$ with $(-\omega,-m)$. It therefore suffices to construct $X$ in the positive-spin sector, suppressing the sector index. The same matrix $X$ applies to both sectors. Throughout this appendix, we use $z=a\cos\theta$, $Y=aT\sin\theta$, and $\epsilon=-\Lambda/3$.

\subsection{Matching the Fourier coefficients} \label{app:X-matching}
Substituting~\eqref{eq:separatedC} into~\eqref{eq:transformedcoupling} gives the equivalent matching condition
\begin{equation}\label{eq:matrix-matching}
 (\det X)(cI-B)=XCX^\dagger \,.
\end{equation}

For the asymptotically flat Kerr--Newman background, Hintz's matrix integrating factor~\cite{Hintz:2026jiw} takes the form
\begin{equation}
 X_{\rm H}=
 \begin{pmatrix}
 \dfrac{Q(\sqrt{\Delta_0}+a\sin\theta)}{f}
 &-\ii\left(M-\dfrac{Q^2}{f}\right)\\[6pt]
 \ii\left(M-\dfrac{Q^2}{g}\right)
 &\dfrac{Q(\sqrt{\Delta_0}-a\sin\theta)}{g}
 \end{pmatrix},
\end{equation}
where $\Delta_0=r^2+a^2-2Mr+Q^2$. With $F=\operatorname{diag}(f,g)$, $z=a\cos\theta$, and $\nu=Mr-Q^2$, we can write it as a sum of radial and angular terms:
\begin{equation}
 FX_{\rm H}
 =\underbrace{Q\sqrt{\Delta_0}\,I+\nu\sigma_2}_{W_r}
 +\underbrace{Mz\sigma_1+Qa\sin\theta\,\sigma_3}_{W_\theta}.
\end{equation}

To extend this structure to KN--(A)dS, we replace $\Delta_0$ by $\Delta$ and $a\sin\theta$ by $Y=aT\sin\theta$, retaining the notation $W_r$ and $W_\theta$ for the resulting blocks. We introduce the following matrix blocks through a common transformation by $i\sigma_1$:
\begin{equation}
\begin{split} \label{eq:construction-blocks}
\mathcal A_\theta &=W_\theta(\ii\sigma_1) =(Mz\sigma_1+QY\sigma_3)(\ii\sigma_1) =\ii Mz\,I-QY\sigma_2  =\begin{pmatrix}
\ii Mz & \ii QY\\
-\ii QY & \ii Mz
\end{pmatrix},\\[4pt]
\mathcal R_r  &=W_r(\ii\sigma_1) =(Q\sqrt{\Delta}\,I+\nu\sigma_2)(\ii\sigma_1) =\nu\sigma_3+\ii Q\sqrt{\Delta}\,\sigma_1  =\begin{pmatrix}
\nu & \ii Q\sqrt{\Delta}\\
\ii Q\sqrt{\Delta} & -\nu
\end{pmatrix}.
\end{split}
\end{equation}
These blocks respectively commute and anticommute with $\sigma_2$. Motivated by this structure, we seek $X$ in the form
\begin{equation}
 X=F^{-1}N \,,\qquad  N=p(r)\mathcal A_\theta+q(\theta)\mathcal R_r \,,
\end{equation}
where $p$ and $q$ are real functions to be determined. The determinant of $N$ therefore contains no cross term proportional to $pq$.

Since a nonzero real scalar factor in $X$ leaves $C$ in Eq.~\eqref{eq:transformedcoupling} unchanged, we first determine the relative weight $\gamma=q/p$ by matching the Fourier coefficients. The terms proportional to $\omega$ and $\Xi m$ in $cI-B$ are $\ii P$, where $P=(K/\sqrt{\Delta})I+\mathcal U\sigma_3$. The $(1,2)$ element of the transformed matrix $\widehat P=(\det X)X^{-1}P(X^\dagger)^{-1}$ is
\begin{equation}
(\widehat P)_{12}=-\frac{2Q(Mz+\ii\nu\gamma)(\gamma K-Y\mathcal U)}{\det(\mathcal A_\theta+\gamma\mathcal R_r)} \,.
\end{equation}
Writing $\gamma K-Y\mathcal U=L_\omega\omega+L_m\Xi m$, we have
\begin{equation}
L_\omega=a^2\sin^2\theta-(r^2+a^2)\gamma \,,\qquad L_m=a(\gamma-1) \,.
\end{equation}
Since $\gamma$ and $\det(\mathcal A_\theta+\gamma\mathcal R_r)$ are real, the $\sigma_1$ and $\sigma_2$ coefficients of each Fourier term are proportional to $MzL_j$ and $-\nu\gamma L_j$, respectively, with $j=\omega,m$. Separation requires the ratios of the $\omega$ to the $\Xi m$ coefficients associated with $\sigma_1$ and $\sigma_2$ to depend only on $\theta$ and $r$, respectively. Where both ratios are defined, they equal $L_\omega/L_m$, which must therefore be independent of both $r$ and $\theta$. Writing $L_\omega/L_m=-(a^2+d)/a$, with $d$ constant, we obtain $\gamma=-(z^2+d)/(r^2-d)$. It follows that $(r^2-d)L_m=-a\Sigma$, so $L_m\ne0$. Using the scalar freedom in $X$, we choose $p=r^2-d$ and $q=-(z^2+d)$, so that
\begin{equation}
N=(r^2-d)\mathcal A_\theta-(z^2+d)\mathcal R_r \,.
\end{equation}
The Fourier combination then simplifies to
\begin{equation}
\gamma K-Y\mathcal U=\frac{\Sigma}{r^2-d}\left[(a^2+d)\omega-a\Xi m\right] \,.
\end{equation}
Substituting this result into the off-diagonal element of $\widehat P$, and using $\det(\mathcal A_\theta+\gamma\mathcal R_r)=\det N/(r^2-d)^2$, gives
\begin{equation}
(\widehat P)_{12}=-2Qa\left[\frac{a^2+d}{a}\omega-\Xi m\right] \times\left(\frac{Mz}{z^2+d}-\ii\frac{\nu}{r^2-d}\right)\frac{\Sigma(r^2-d)(z^2+d)}{\det N} \,.
\end{equation}
Here $Mz/(z^2+d)$ depends only on $\theta$, while $\nu/(r^2-d)$ depends only on $r$. Define $\mathcal F(r,\theta)=\Sigma(r^2-d)(z^2+d)/\det N$. The $\sigma_1$ and $\sigma_2$ Fourier coefficients are then proportional to $[Mz/(z^2+d)]\mathcal F$ and $[\nu/(r^2-d)]\mathcal F$, respectively. Since the former must depend only on $\theta$ and the latter only on $r$, separation requires $\partial_r\mathcal F=\partial_\theta\mathcal F=0$. Thus $\mathcal F$ must be constant, implying $\det N\propto\Sigma(r^2-d)(z^2+d)$.

From the explicit forms of $\mathcal A_\theta$ and $\mathcal R_r$ in Eq.~\eqref{eq:construction-blocks}, we obtain
\begin{equation}
\det N=-(r^2-d)^2(M^2z^2+Q^2Y^2)-(z^2+d)^2(\nu^2-Q^2\Delta) \,.
\end{equation}
For $\det N$ to be proportional to $\Sigma(r^2-d)(z^2+d)$, the radial polynomial $\nu^2-Q^2\Delta$ must contain the factor $r^2-d$. Since this polynomial is at most quadratic in $r^2$, with the coefficient of $r^4$ given by $-\epsilon Q^2$, we write $\nu^2-Q^2\Delta=(r^2-d)(-\epsilon Q^2r^2+\beta)$. Comparing coefficients and writing $d=a^2\eta$, we obtain
\begin{equation}
M^2\eta=Q^2(1+\eta)(1+\epsilon a^2\eta) \,.
\end{equation}
We choose the root $\eta$ in Eq.~\eqref{eq:alg-parameters}, which has a smooth asymptotically flat limit. Defining $A_r=r^2-d$ and $D_\theta=z^2+d$, we obtain $N=A_r\mathcal A_\theta-D_\theta\mathcal R_r$. The two polynomials appearing in $\det N$ then factor as
\begin{equation}\label{eq:alg-factorizations}
\nu^2-Q^2\Delta=\frac{Q^2}{\eta}A_r(1-\epsilon\eta r^2) \,,\qquad
M^2z^2+Q^2Y^2=\frac{Q^2}{\eta}D_\theta(1+\epsilon\eta z^2) \,.
\end{equation}
Using $\kappa=Q^2(1-\epsilon a^2\eta^2)/\eta$, we obtain
\begin{equation}
\det N=-\kappa A_rD_\theta\Sigma \,.
\end{equation}

Fixing the overall scalar normalization, we obtain
\begin{equation}\label{eq:construction-compact-X}
X=\frac1{\kappa Q\sqrt{A_rD_\theta}}\operatorname{diag}(f^{-1},g^{-1})N \,.
\end{equation}
This reproduces Eq.~\eqref{eq:alg-frame}. To complete the algebraic matching, we substitute this matrix into Eq.~\eqref{eq:matrix-matching} and verify that the terms independent of $\omega$ and $m$ also separate. Together with the Fourier terms determined above, this yields all four coefficients in Eq.~\eqref{eq:alg-separated-coefficients}.

Finally, we verify the derivative terms arising from the transformation by $X$. Using $\Dop\log[(A_rD_\theta)^{-1/2}]=-r\sqrt{\Delta}/A_r-\ii zY/D_\theta$, direct differentiation gives
\begin{equation}\label{eq:construction-detX}
X^{-1}(\Dop X-AX)=(h_r+h_\theta)\sigma_2,\qquad \det X=-\frac{1}{\kappa Q^2} \,.
\end{equation}
Since $A$ is Hermitian, $h_r+h_\theta$ is real, and $\det X$ is constant, taking the Hermitian conjugate gives the corresponding identity for the second transformed equation. Together with the algebraic matching, this establishes Eq.~\eqref{eq:alg-chiral}.

Note that the absence of an identity term in Eq.~\eqref{eq:construction-detX} follows from our choice of normalization and is not necessary for separation. Under the scalar freedom in Eq.~\eqref{eq:freedom}, we may take $\widetilde X=s_r(r)s_\theta(\theta)X$, where $s_r$ and $s_\theta$ are positive smooth functions. The derivative identity then becomes
\begin{equation}
\widetilde X^{-1}(\Dop\widetilde X-A\widetilde X) =(h_r+h_\theta)\sigma_2+ \left(\sqrt{\Delta}\,\partial_r\log s_r-\ii T\,\partial_\theta\log s_\theta\right)I \,.
\end{equation}
The additional identity term separates into radial and angular parts, while the matrix $C$ remains unchanged. For example, the unnormalized choice $\widetilde X=F^{-1}N$ has an identity coefficient $r\sqrt{\Delta}/A_r+\ii zY/D_\theta$, which is canceled by the factor $(A_rD_\theta)^{-1/2}$ in Eq.~\eqref{eq:construction-compact-X}.

\subsection{Transport interpretation of the separating matrix}
The derivative identity~\eqref{eq:construction-detX} implies $\Dop X=AX+(h_r+h_\theta)X\sigma_2$. Consequently, the matrix $J_X=X\sigma_2X^{-1}$ satisfies the transport equation
\begin{equation}
\Dop J_X=[A,J_X] \,, \qquad J_X^2=I \,.
\end{equation}
Let $v_s=(1,s\ii)^T$, with $s=\pm1$, be the eigenvectors of $\sigma_2$, satisfying $\sigma_2v_s=sv_s$. It follows that $J_X(Xv_s)=sXv_s$. This suggests recovering $X$, up to normalization, from the eigenvectors of a matrix satisfying the transport equation. Gray and Ovcharenko developed an eigendirection-based construction for Pleba\'nski--Demia\'nski spacetimes, including the KN--(A)dS limit~\cite{Gray:2026rzu}. Here we instead obtain the transported matrix by adding an explicit correction to a seed constructed from Hintz's integrating factor.

Specifically, replacing $\Delta_0$ with $\Delta$ and $a\sin\theta$ with $Y=aT\sin\theta$ in Hintz's integrating factor~\cite{Hintz:2026jiw}, we obtain
\begin{equation}
\Dop Z-AZ=\epsilon Q\Sigma I \,, \qquad Z=F^{-1}(Q\sqrt{\Delta}\,I+Mz\sigma_1+\nu\sigma_2+QY\sigma_3) \,.
\end{equation}
For $\epsilon=0$, the matrix $J_0=-Z\sigma_2\adj Z$ satisfies the transport equation. For nonzero $\epsilon$, $J_0$ no longer satisfies the transport equation exactly. Defining $V=\begin{pmatrix}0&-\ii g^2\\ \ii f^2&0\end{pmatrix}$, we find
\begin{equation}
\begin{aligned}
\Dop J_0-[A,J_0]&=-\epsilon Q\Sigma(\sigma_2\adj Z+Z\sigma_2) \,,\\
\Dop V-[A,V]&=\frac{\Sigma}{Q}(\sigma_2\adj Z+Z\sigma_2) \,.
\end{aligned}
\end{equation}
Adding the correction $\epsilon Q^2V$ restores the transport equation, yielding
\begin{equation}
J =-Z\sigma_2\adj Z+\epsilon Q^2V \,,\qquad \Dop J =[A,J] \,, \quad \tr J=0 \,,\quad J^2=\kappa^2I \,.
\end{equation}
Here $\kappa$ is the positive constant defined in Eq.~\eqref{eq:alg-parameters}.

To recover $X$, we project onto the two eigenspaces of $J$ using $\Pi_s=(I+sJ/\kappa)/2$. With $b_s=\ii Mz-sQY$ and $d=a^2\eta$, direct calculation using~\eqref{eq:alg-factorizations} gives
\begin{equation}
F\Pi_sF^{-1}v_s=\frac{\bar b_s}{\kappa\Sigma D_\theta}\big[A_rb_sv_s-D_\theta(\nu-sQ\sqrt{\Delta})v_{-s}\big] \,.
\end{equation}
Since $\mathcal A_\theta v_s=b_sv_s$ and $\mathcal R_rv_s=(\nu-sQ\sqrt{\Delta})v_{-s}$, the bracketed expression is precisely $Nv_s$. Moreover, $|b_s|^2=M^2z^2+Q^2Y^2>0$ and $\kappa\Sigma D_\theta\ne0$ throughout the black-hole exterior, so the prefactor is nonzero. The projected vectors are therefore proportional to $F^{-1}Nv_s$. After rescaling these vectors and fixing the normalization as in Eq.~\eqref{eq:construction-compact-X}, we recover the separating matrix $X$ obtained by Fourier matching. Consequently,
\begin{equation}
J=\kappa X\sigma_2X^{-1} \,.
\end{equation}
%

\section{Invertibility of \texorpdfstring{$X$}{X} in the black hole exterior} \label{app:invertibility}

From~\eqref{eq:construction-compact-X} and \eqref{eq:construction-detX}, smooth invertibility reduces to showing that $\kappa$, $A_r$, $D_\theta$, and $\Sigma=fg$ do not vanish in the physical regions. Since $\eta>0$, we have $D_\theta>0$, while $\Sigma>0$ follows directly from the background geometry.

For KN--AdS, the explicit form of $\Delta$, together with $\Delta(r_+)=0$ and $\Delta'(r_+)>0$, implies
\begin{equation}
 M>|Q|\left(1+\frac{|a|}{L}\right) \,.
\end{equation}
It follows that $\kappa>0$, $\eta>0$, and $|a|\eta/L<1$. Evaluating~\eqref{eq:alg-factorizations} at $r=r_+$ gives
\begin{equation}  \label{eq:ads-horizon-factorization}
 (Mr_+-Q^2)^2 =\frac{Q^2}{\eta}  (r_+^2-\eta a^2)  \left(1-\frac{\eta r_+^2}{L^2}\right) \,.
\end{equation}
Suppose that $r_+^2\leq\eta a^2$. Then $1-\frac{\eta r_+^2}{L^2}\geq 1-\frac{\eta^2a^2}{L^2}>0$. The right-hand side of~\eqref{eq:ads-horizon-factorization} is therefore nonpositive, whereas the left-hand side is nonnegative. Hence both sides must vanish, forcing $r_+^2=\eta a^2$ and $Mr_+=Q^2$. Differentiating the first identity in~\eqref{eq:alg-factorizations} at $A_r=\nu=0$ gives $\Delta'(r_+)=-2r_+(\eta^{-1}-r_+^2/L^2)<0$, contrary to the simple outer-horizon condition $\Delta'(r_+)>0$. Hence $r_+^2>\eta a^2$, and therefore $A_r>0$ throughout $r>r_+$.

For KN--dS, the definitions give $\kappa>0$ and $\eta>0$. At the event horizon $r=r_b$, we have similarly
\begin{equation}
(Mr_b-Q^2)^2 =\frac{Q^2}{\eta}  (r_b^2-\eta a^2) \left(1+\frac{\eta r_b^2}{L^2}\right) \,.
\end{equation}
The last factor is positive, so $r_b^2\geq\eta a^2$. Equality would again imply $\Delta'(r_b)<0$, contradicting $\Delta'(r_b)>0$. Thus $A_r>0$ throughout $r_b<r<r_c$.

For $0<\theta<\pi$, the matrix $X$ is smooth outside the event horizon ($r>r_+$) in KN--AdS and between the event and cosmological horizons ($r_b<r<r_c$) in KN--dS. Since $\det X\ne0$ by Eq.~\eqref{eq:construction-detX}, the field transformation \eqref{eq:change} has a smooth inverse throughout these regions in both spin sectors.

\section{Axis and horizon conditions} \label{app:physical-boundaries}
The four fields in each sector have spin weights $(2\varsigma,\varsigma,\varsigma,2\varsigma)$. Near the north and south axes, smooth tetrads are obtained by $m^\mu\mapsto e^{\ii\phi}m^\mu$ and $m^\mu\mapsto e^{-\ii\phi}m^\mu$, respectively. Thus, for a spin-$s$ mode proportional to $e^{\ii m\phi}$, smoothness requires $t^{|m+s|}A_N(t^2)$ and $t^{|m-s|}A_S(t^2)$ at the north and south axes, where $A_N$ and $A_S$ are smooth functions.

Write
\begin{equation}
X=\mathcal P+\sqrt\Delta\,\mathcal B \,,\quad \mathcal B=-\frac{\ii D_\theta}{\kappa\sqrt{A_rD_\theta}} \operatorname{diag}(f^{-1},g^{-1})\sigma_1 \,,
\end{equation}
where $\mathcal P$ is independent of $\sqrt\Delta$. Since $\adj(X^\dagger)=-\mathcal P+\sqrt\Delta\,\mathcal B$, the inverse transformation gives
\begin{equation}
\begin{aligned} \label{eq:boundary-exact-inverse}
Y_s:=y_++y_- &=\frac{(\sqrt\Delta\,\mathcal B u_1 +\ii\mathcal P\sigma_2u_2)\chi}{\det X} \,, \\
Y_d:=y_+-y_- &=\frac{(\mathcal P u_1 +\ii\sqrt\Delta\,\mathcal B\sigma_2u_2)\chi}{\det X} \,.
\end{aligned}
\end{equation}
Here the factor $(A_rD_\theta)^{-1/2}$ is smooth and nonzero at the axes and horizons. At either axis, the diagonal elements of $\mathcal P$ are even in $t$, its off-diagonal elements are odd, and the off-diagonal elements of $\mathcal B$ are even. Together with~\eqref{eq:physical-axis-domain}, these parity properties give the required spin-weighted behavior of the radiative NP variables through~\eqref{eq:chiralvars}. Conversely, smoothness of all four radiative variables requires $y_\pm=O(t^{1/2})$. The invertible transformation~\eqref{eq:change} and the ansatz~\eqref{eq:ansatz} then imply $\chi=O(t^{1/2})$ for any nontrivial separated solution, which excludes the singular angular branch.

For a simple horizon, define
\begin{equation}
 r_*'=\frac{r^2+a^2}{\Delta} \,, \qquad   \Phi_*'=\frac{a\Xi}{\Delta} \,.
\end{equation}
At a future event horizon, the ingoing coordinates $v=t+r_*$ and $\phi_{\rm in}=\phi+\Phi_*$ give the factor $H^{-1}$. The tetrad $(\Delta\ell,\Delta^{-1}n,m,\bar m)$ is smooth there, so the following combinations of radiative variables must also be smooth:
\begin{equation}
H^{-1}(\Delta^2\psi_2,\Delta\psi_1,\Delta^2k,\Delta s) \,,\qquad  H^{-1}(\Delta^{-2}\psi_{-2},\Delta^{-1}\psi_{-1}, \Delta^{-2}k_-,\Delta^{-1}s_-) \,.
\end{equation}
At a future cosmological horizon, the outgoing coordinates $u=t-r_*$ and $\phi_{\rm out}=\phi-\Phi_*$ give the factor $H$, and the original tetrad is smooth.

Substituting~\eqref{eq:boundary-exact-inverse} into~\eqref{eq:chiralvars} and including the horizon factors above gives~\eqref{eq:physical-horizon-data}. After the powers of $H$ and $\Delta$ are removed, the remaining transformation has a nonsingular horizon limit: at $\Delta=0$, $\mathcal P=X$ and $\det\mathcal P=\det X\neq0$. Thus~\eqref{eq:physical-horizon-data} is precisely the condition that the radiative variables are smooth at the horizon in the regular tetrad. The same even--odd argument applies where an axis meets a horizon.

\section{Asymptotically flat limit} \label{app:flat-equivalence}
Let $M,a,Q$ be fixed, with $M^2>a^2+Q^2$ and $aQ\ne0$. As $L\to\infty$, the parameters satisfy $\kappa\to M^2-Q^2$ and $\eta\to Q^2/(M^2-Q^2)$. Taking this limit in~\eqref{eq:alg-frame}, we find the following relation to Hintz's matrix integrating factor $X_H$~\cite{Hintz:2026jiw}:
\begin{equation}
X_H=-QX \, \frac{\nu I+Q\sqrt{\Delta}\,\sigma_2}{\sqrt{A_r}} \times \frac{Qa\sin\theta\,I+\ii Ma\cos\theta\,\sigma_2}{\sqrt{D_\theta}} \,.
\end{equation}
The two factors to the right of $X$ depend only on $r$ and $\theta$, respectively, and are smooth and invertible throughout the subextremal KN exterior $r>r_+$. Thus, in the asymptotically flat limit, $X$ and Hintz's integrating factor are equivalent under the frame freedom~\eqref{eq:freedom}.

\bibliographystyle{JHEP}
\bibliography{refs}
\end{document}